\documentclass[%
reprint,  
superscriptaddress,
nofootinbib,
amsmath,amssymb,
aps,
pra,
raggedbottom,
onecolumn,
xcolor,
]{revtex4-1}

\usepackage{graphicx}
\usepackage{dcolumn}
\usepackage{bm}
\usepackage{lipsum}
\usepackage{float}
\usepackage{subfigure}

\usepackage[colorlinks=true,bookmarks=false]{hyperref}
\hypersetup{linkcolor=blue,citecolor=red,urlcolor=blue}   

\begin{document}


\title{Segregation of charged particles in shear induced diffusion}

\author{R. Yoshimatsu} \email{ryutay@phys.ethz.ch} 
\affiliation{%
Computational Physics for Engineering Materials, IfB, ETH Zurich, Wolfgang-Pauli-Strasse 27, 8093 Zurich, Switzerland
}

\author{N.A.M. Ara\'ujo}
\affiliation{%
Departamento de F\'isica, Faculdade de Ci\'encias, Universidade de Lisboa, P-1749-016 Lisboa, Portugal, and Centro de F\'isica Te\'orica e Computacional, Universidade de Lisboa, P-1749-016 Lisboa, Portugal
}

\author{T. Shinbrot}
\affiliation{%
Department of Biomedical Engineering, Rutgers University, Piscataway, New Jersey, 08854, USA 
}

\author{H.J. Herrmann}
\affiliation{%
Computational Physics for Engineering Materials, IfB, ETH Zurich, Wolfgang-Pauli-Strasse 27, 8093 Zurich, Switzerland
}
\affiliation{%
Departamento de F\'isica, Universidade Federal do Cear\'a, 60451-970 Fortaleza,Cear\'a, Brazil
}

                  
\date{May 10, 2017}
\begin{abstract}
We study segregation of a binary mixture of similarly charged particles under shear using particle-based simulations. We simulate particle dynamics using a discrete-element model including electrostatic interactions and find that particles segregate according to their net charge. Particles that are charged twice as strong as other particles of the same electrical sign are seen more at insulating boundaries with which we shear the system. Weakly charged particles, on the other hand, stay more in the center of the sheared bed. We propose a simple model based on electrostatic potential energy to understand this segregation. The model shows that the segregated system we observe in our simulations is indeed the most favorable configuration in terms of electrostatic potential energy. Our simulations further show that for a given packing fraction there is an optimal shear velocity where the segregation maximally intensifies. We show that this maximum results from a competition between diffusional and Coulomb fluxes. For a larger shear velocity, diffusion suppresses segregation.
\end{abstract}

\maketitle

\section{\label{introduction}Introduction}
Grains have long been known to segregate \cite{aranson2006patterns}, or de-mix, based on size \cite{duran1993arching, knight1993vibration, zik1994rotationally, pouliquen1997fingering, hill2008isolating, harrington2013suppression, van2015underlying}, shape \cite{abreu2003influence, kyrylyuk2011isochoric, shimoska2013effect}, density \cite{shinbrot1998reverse, fan2015shear}, friction coefficient \cite{gillemot2017shear}, or other material properties \cite{windows2014effects}. De-mixing is commonly seen both in natural processes, for example in segregation of geological debris \cite{felix2004relation, kleinhans2004sorting, johnson2012grain, kokelaar2014fine, haas2015effects}, and in industrial systems, for example in pharmaceutical mixing where uniformity is crucial to producing safe and effective medicines \cite{makse1998spontaneous, ottino2000mixing, shinbrot2000nonequilibrium}.

Segregation based on particle charge has also been widely applied in industry. For example, in plastics recycling, differential charging is used to separate different types of plastics \cite{inculet1998electrostatic}. In electrospraying and xerography, electrostatic forces adhere and localize charged grains \cite{duke2002surface, mehrotra2007spontaneous}. In a biological context, cell charging is used to separate dead from living and healthy from cancerous cells by applying non-uniform electric fields \cite{martinez2012microfabrication}. Additionally, it has been shown that mechanically identical grains that are charged with two different magnitudes of the same sign spontaneously segregate while sliding down a chute \cite{mehrotra2007spontaneous} and another study has found that charged particles in a fluidized granular bed migrate toward the walls when all particles are charged equally \cite{kolehmainen2016hybrid}. Recently, it has also been shown that a binary mixture of mechanically dissimilar particles otherwise known to segregate can be made to mix by charging particles with opposite electrical signs \cite{schella2017charging}. 

Despite the prevalence and practical importance of particle segregation, comparatively little is known about the effects of particle charge under controlled circumstances. In the present work, we study the segregation of mechanically identical but charged particles in a shear flow using particle-based simulation, with the goal of understanding the fundamental mechanics underlying segregation of charged particles. We use discrete-element method (DEM) \cite{poschel2005computational} simulations and demonstrate that a granular flow with a binary mixture of similarly charged particles segregates under shear according to a precise formula.

The paper is structured as the following: in section 2, we describe the simulation, then in section 3, we analyze results, and in section 4, we draw conclusions.

\begin{figure}
\centering
\includegraphics[width=8cm]{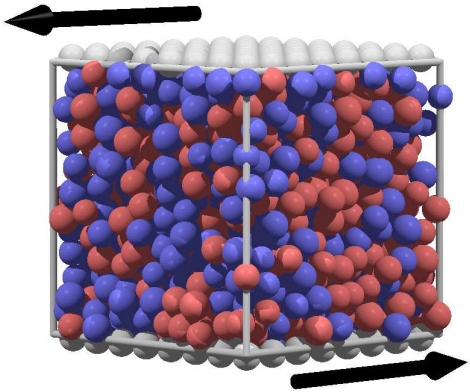}
\caption{Snapshot of the discrete-element method (DEM) simulation. Red particles are electrically charged twice as strong as blue particles with the same sign. Grey particles are electrically neutral and embedded into the plates. Arrows show the directions of the shear.}
\label{profile}
\end{figure}

\section{\label{model}Model}
We perform DEM simulations of spherical dielectric particles in three dimensions, as sketched in Fig.\,\ref{profile}. The grain sizes are polydisperse to prevent crystallization: the radius of each grain is normally distributed around the mean radius, $\bar{R}=0.5$ mm, with standard deviation 5\%. We are interested in electrostatic effects between mechanically identical grains, so all grains have the same mass: $m=\frac{4}{3} \pi \bar{R}^3 \rho_{g}$, where $\rho_g$ = 2.4 g/cm$^3$ is the density of glass. We have confirmed, using our simulation, that the polydipersity that we introduce produces no segregation on its own. 

We use the standard procedures to evaluate forces and torques on interacting particles \cite{yoshimatsu2016field}. In short, for the elastic force, we use the model of Walton and Braun \cite{walton1986viscosity} with two elastic coefficients, $k_{l}$ = 0.07 and $k_{u}$ = 0.08, to achieve the restitution coefficient, $\sqrt{k_l / k_u}$ = 0.935. For friction, we use the standard kinetic friction model with kinetic friction coefficient, $\mu_k$ = 0.4. We compute Coulomb forces directly using a virtual charge fixed at the center of each grain. Long-range electrostatic forces between grains are computed using the Particle-Particle Particle-Mesh Method described elsewhere \cite{hockney1988computer, yoshimatsu2016field}. Unlike prior works \cite{yoshimatsu2016field, yoshimatsu2017self}, dipole moments are neglected, and we compute electrostatic forces only using grains' net charges. The time step for the numerical integration is 50 msec., which produces more than 100 steps per collision for the fastest moving grains. The boundaries in horizontal directions are periodic with respect to both mechanical and electrical forces.

To obtain an initial configuration with a desired solid fraction, $\phi$, we place 10$^3$ grains randomly in a three dimensional box with stiff boundaries. The top and bottom boundaries are 20$\bar{R}$ $\times$ 20$\bar{R}$ in size, into which we embed neutral insulating grains of radius $\bar{R}$ on a fixed grid. We wait until the initial packing relaxes and grain velocities become negligibly small under gravity. We then turn off the gravity and compress the system vertically by lowering the top boundary with a constant slow speed. Once a desired solid fraction, $\phi$, is achieved, we stop the compression. Grains are all electrically neutral during the initialization process and gravity is always turned off thereafter.

Fixed boundary conditions are implemented at the top and bottom. We shear the packing by moving the top and bottom boundaries horizontally in opposite directions with a fixed speed, $v_s$ (in units of 2$\bar{R}$/sec.). We apply shear until the system reaches a steady velocity field, then we distribute charges on all non-boundary particles with values randomly chosen to be either +$q$ or +2$q$, where $q$ $\sim$ 1.8$\times$10 pC. We set this time to be $t$ = 0 in the results that follow. Note that the transition from an unjammed to a jammed state occurs at $\phi$ $>$ 0.6, below which the packing remains collisional even for a  small shear velocity \cite{o2003jamming}. 

\section{\label{results}Results}
\begin{figure}
\centering
\includegraphics[width=11cm]{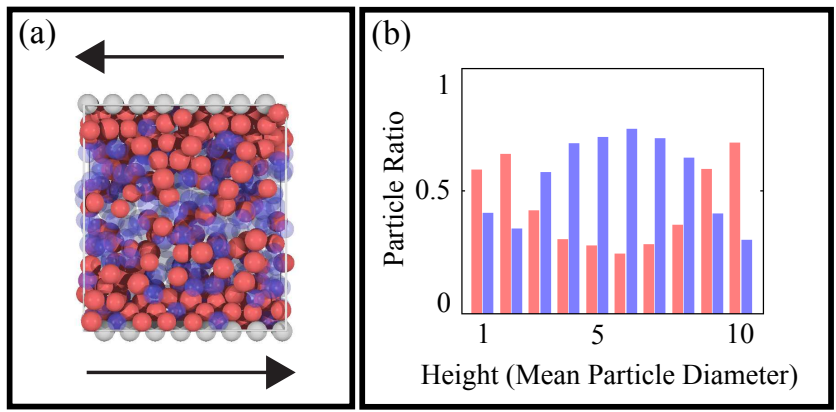}
\caption{(a) Snapshot of charged particles in sheared bed at $t$ = 500 sec. for $\phi=0.55$ and $v_{s}=10$. Weakly charged particles (blue) are displayed transparently to demonstrate the separation of particles. (b) Histograms of the numbers of strongly charged particles (red) and weakly charged particles (blue) as a function of height for $\phi=0.55$ and $v_{s}=10$. We divide the system into 10 horizontal slices of thickness $2\bar{R}$ and count the number of red and blue particles every 50 seconds after the system reached an asymptotic steady state. We then average these measurements over 5 configurations and compute the ratio between red and blue particles.}
\label{0}
\end{figure}

Figure \,\ref{0} (a) shows a snapshot of a simulation at $t$ = 500 sec. for packing fraction $\phi$ = 0.55 and shear velocity $v_s$ = 10 (in units of 2$\bar{R}$/sec.). Qualitatively, more strongly charged particles (red) gather closer to the shearing boundaries than more weakly charged particles (transparent blue):  this is quantified in Fig.\,\ref{0} (b), where for the simulation shown in Fig.\,\ref{0} (a), we plot a histogram of the mean number of strongly (red) and weakly (blue) charged particles as a function of height (details in figure caption).

We evaluate the degree of segregation in our sheared bed using the segregation parameter, $S$, defined as: 
\begin{equation}
S=\frac{1}{N}\sum_{i=1}^{N} \frac{(n_{i,e} - n_{i,d})^2}{(n_{i,e} + n_{i,d})^2},
\label{segregation}
\end{equation}
where $N$ is the total number of particles, $n_{i,e}$ is the number of like-charged particles around the i$^{th}$ particle within the cutoff radius, $r_{cut}=3\bar{R}$, from the center of that particle, and $n_{i,d}$ is the number of differently-charged particles within the same area. For a perfectly segregated system, $S$ = 1, because particles only have like-charged neighbors, and for a perfectly mixed case, $S$ = 0, because the number of similar and dissimilar neighbors is equal. 

\begin{figure}
\centering
\includegraphics[width=15cm]{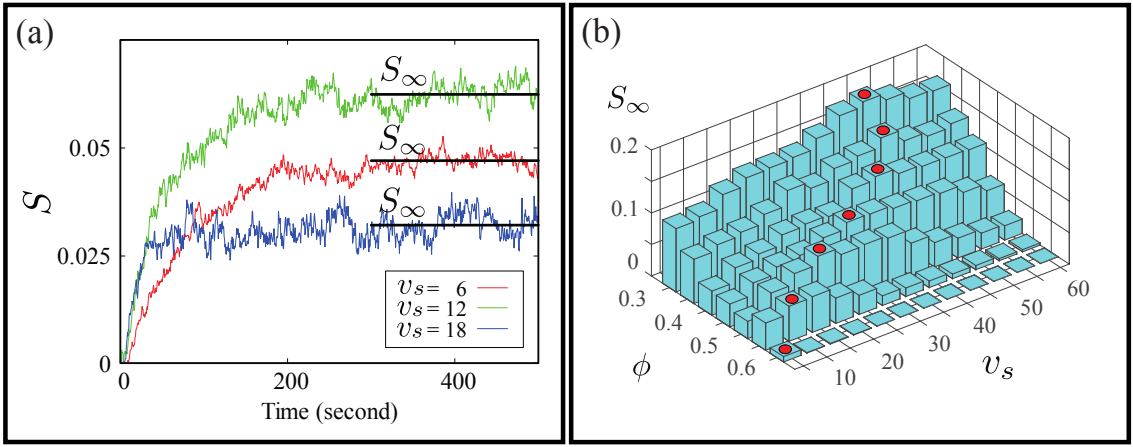}
\caption{(a) Time evolution of the segregation parameters, $S$, with a fixed packing fraction, $\phi = 0.55$, and varying shear velocities, $v_s$. Each curve is an average over five independent simulations using the same parameters but different initial configurations. Black lines show the mean values, $S_{\infty}$, calculated after $t=300$ sec., a time where we assume $S$ appears to have reached its asymptotic value. (b) Phase diagram showing $S_{\infty}$ for different $\phi$ and $v_s$. Red spots correspond to the maximum $S_{\infty}$ values for a given $\phi$. For $\phi$ $>$ 0.6, the bed attains a nearly jammed state, and segregation does not occur.}
\label{1}
\end{figure}

Figure \,\ref{1} (a) shows the time evolution of $S$ for fixed $\phi$ = 0.55 and several shear velocities $v_s$. In all cases, $S$ grows with time until it saturates at $S_{\infty}$, but surprisingly $S_{\infty}$ depends non-monotonically on $v_s$: i.e. sheared beds segregate maximally at moderate shearing velocities. Similar non-monotonic behaviors are seen for different packing fractions, as shown in Fig.\,\ref{1} (b). Apparently segregation produced by dissimilar charges exhibits nontrivial dynamics, which we explore next.

\begin{figure}
\centering
\includegraphics[width=17.75cm]{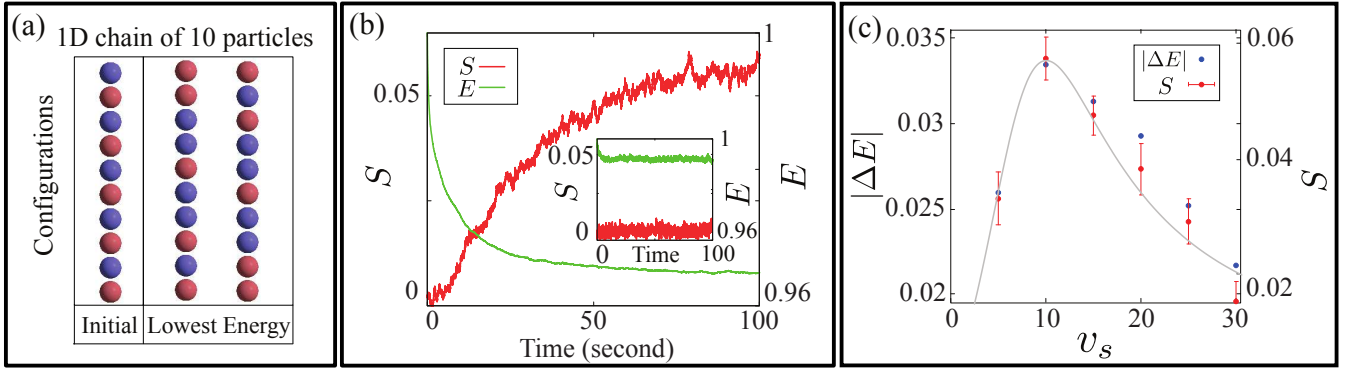}
\caption{(a) Cartoon showing the initial fully mixed configuration and two configurations with the lowest electrostatic potential energy of a one-dimensional chain of ten charged particles. Red particles are charged +2$q$ and blue particles +$q$, and the separation distance between particles is fixed. (b) Time evolution of $S$ (red) and total electrostatic potential energy, $E$ (green), defined as: $E$=$\frac{k_e}{2}\sum_{i \neq j}^{N} \frac{q_{i}q_{j}}{r_{ij}}$, where $r_{ij}$ is the distance between the particles $i$ and $j$, for $\phi=0.55$ and $v_{s}=10$. $E$ is normalized by its initial value, $E_{ini}$. Inset shows the same measurements for $\phi=0.55$ and $v_s=60$, where $S$ does not grow much. (c) $\mid$$\Delta E$$\mid$ (blue) and $S_{\infty}$ (red) versus $v_s$ for $\phi$ = 0.55. $\mid$$\Delta E$$\mid$ is the absolute difference in $E$ between the initial and the final configurations: $\mid$$\Delta E$$\mid$=$\mid$$E_{ini}$-$E_{\infty}$$\mid$, where $E_{\infty}$ is $E$ in an asymptotic steady state. Grey curve to aid the eye is the same shown in Fig.\,\ref{3} (c), rescaled.}
\label{2}
\end{figure}

To understand the segregation observed, we note that since all charges are of the same sign, the electrostatic energy of the particle bed is minimized when the most highly charged particles are as far away from other charges as possible. This is a simple consequence of the fact that the self-energy of a charge density, $\rho(\vec{x})$, is: $E$ = $\frac{k_{e}}{2} \int \int \frac{\rho(\vec{x}) \rho(\vec{x'})}{\mid \vec{x}-\vec{x'} \mid} dx^3 dx'^3$ \cite{jackson1975electrodynamics}, where $k_e$ is Coulomb's constant, hence locating the largest charges along boundaries, beside which $\rho(\vec{x})$ = 0, minimizes the energy. In our simulations, the boundary particles are insulating, however, we remark that for conducting boundaries, image charges would be of opposite sign and so there would be even stronger energetic advantage when locating the highest charges near the boundaries \cite{mehrani2017overview}. To study if such spatial arrangement of particles actually minimizes the electrostatic potential energy of the system, we construct a simple model consisting of ten charged particles forming a one-dimensional chain. Half of the particles are strongly charged and the other half are weakly charged like in our simulations. Depending on the ordering of the particles, the chain takes different electrostatic potential energies, which we compute as: $E$ = $\frac{k_{e}}{2} \sum^{10}_{i \neq j} \frac{q_{i} q_{j}}{r_{ij}}$, where $q_i$  and $q_j$ are the charges of the i$^{th}$ and j$^{th}$ grains, and $r_{ij}$ is the distance between the centers of the two grains. A comparison of energies of different configurations shows that the most favorable configurations have strongly charged particles right next to boundaries and more weakly charged particles in the center as shown in Fig.\,\ref{2} (a). Notice that the model does not exhibit a perfect de-mixing of strongly and weakly charged particles, since the lowest energy configurations also contain some strongly charged particles in the center.

To examine if the same argument holds for our sheared system, we fix $\phi$ = 0.55 and measure the time evolution of the total electrostatic potential energy computed the same way. As shown in Fig.\,\ref{2} (b), $E$ decreases as $S$ grows for a system that strongly segregates with $v_s$ = 10 (main plot), and stays almost unchanged for a system that does not segregate with $v_s$ = 60 (inset). Furthermore, the drop in the electrostatic potential energy: $\mid$$\Delta E$$\mid$=$\mid$$E_{ini}$-$E_{\infty}$$\mid$, where $E_{ini}$ is the energy at $t$=0 and $E_{\infty}$ is the energy in an asymptotic steady state, behaves similarly to $S_{\infty}$. This is shown in Fig.\,\ref{2} (c), where we plot $\mid$$\Delta E$$\mid$ as a function of $v_s$ for $\phi$ = 0.55. Note that for $\phi$ = 0.55, $S_{\infty}$ takes its maximum at $v_s$ = 10, in agreement with Fig.\,\ref{1} (b), which shows that the maximally segregated system has the lowest electrostatic potential energy. We have measured the mean inter-particle distance: $<$$r_d$$>$ = $\frac{1}{N}\sum_{i=1}^{N} r_{ik}$, where $r_{ik}$ is the distance between the centers of particle $i$ and its closest neighbor particle $k$, and we confirm that this remains unchanged for an electrically neutral and a charged system. This implies that the packing does not expand due to charging and so the cause of the drop in the electrostatic potential energy is the rearrangement -i.e. the segregation- of particles. Moreover, we have also employed Lees-Edwards boundary conditions \cite{lees1972computer}, which connect the lower and the upper boundaries via a periodic boundary. In this case, the energetic advantage of placing highly charged particles near the boundaries is removed, and segregation does not occur.

\begin{figure}
\centering
\includegraphics[width=15cm]{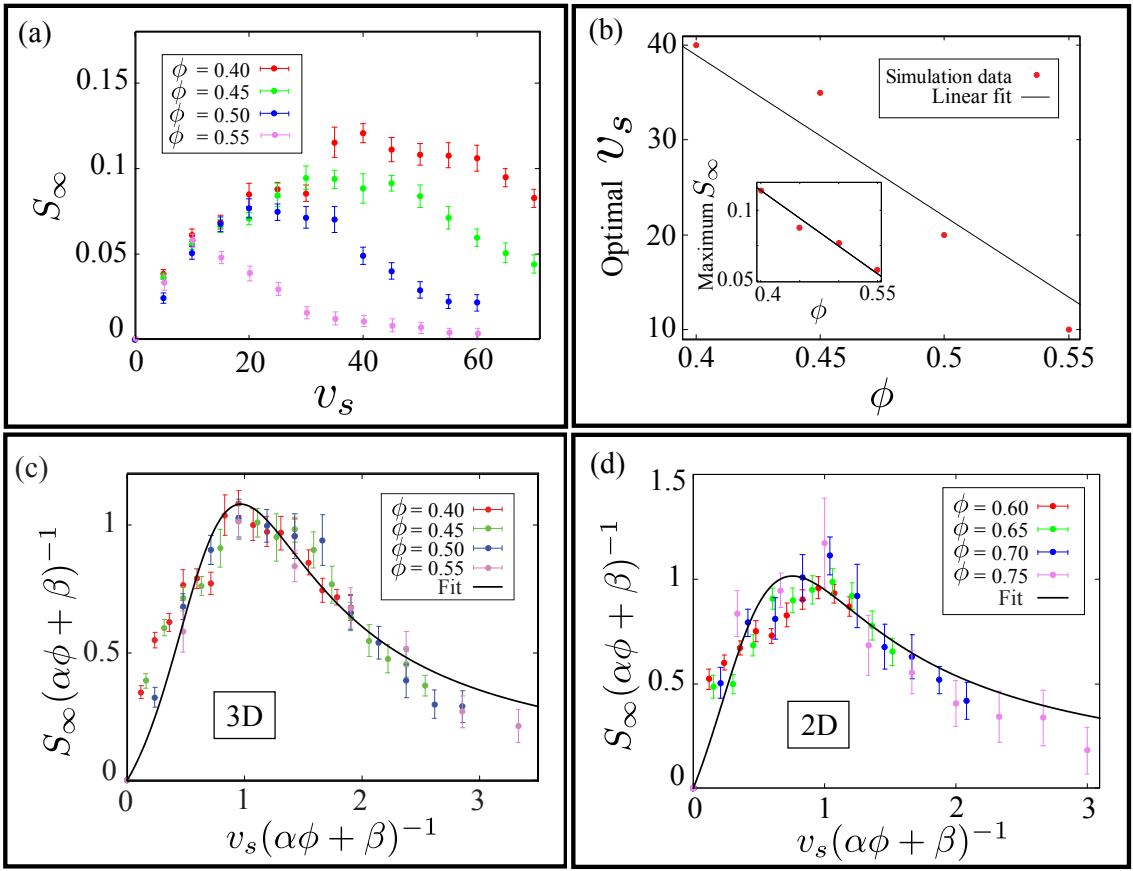}
\caption{(a) $S_{\infty}$ versus $v_s$ for several choices of $\phi$. Error bars indicate the standard deviation (over 20 trials). (b) Optimal $v_s$ versus $\phi$ and a linear least-squares fit. Inset shows maximum $S_{\infty}$ versus $\phi$ with a linear fit obtained in the same way. (c) $S_{\infty}$ versus $v_s$, scaled using linear functions of $\phi$: $\alpha \phi+\beta$, where for the vertical axis, ($\alpha$, $\beta$) = (-0.4, 0.28), and for the horizontal axis, ($\alpha$, $\beta$) = (-170, 107). Black curve is a fit to Eq.\,\eqref{model}. (d) Scaled $S_{\infty}$ versus scaled $v_s$ from two dimensional DEM simulation. The axes are scaled by linear functions as before, where for the vertical axis ($\alpha$, $\beta$) = (-0.53, 0.45) and for the horizontal axis ($\alpha$, $\beta$) = (-360, 300). Again the black curve is a fit to Eq.\,\eqref{model}.}
\label{3}
\end{figure}

Next, we discuss the non-monotonic dependency of $S_{\infty}$ on $v_s$. Figure \,\ref{3} (a) shows $S_{\infty}$ versus $v_s$ for several $\phi$. Each curve has its distinct maximum $S_{\infty}$ and optimal $v_s$, a shear velocity where $S_{\infty}$ maximizes. Both the maximum $S_{\infty}$ and the optimal $v_s$ decrease with $\phi$. To further investigate the relationship between the packing fraction and the segregation, we note that the maximum $S_{\infty}$ and the optimal $v_s$ vary nearly linearly with $\phi$, as shown in Fig.\,\ref{3} (b).  Consequently, we scale $S_{\infty}$ and $v_s$ using simple linear functions, $\phi$: $\alpha \phi+\beta$, where $\alpha$ and $\beta$ are obtained from least-squares linear fits of the maximum $S_{\infty}$ vs. $\phi$ (for the vertical axis), and the optimal $v_s$ vs. $\phi$ (for the horizontal axis). After the scaling, curves collapse onto a single curve with a peak close to unity, in keeping with the observation that the maximum segregation and the optimal shear velocity both decrease linearly with the packing fraction, as shown in Fig.\,\ref{3} (c) . We also did simulations of two dimensional systems. Figure \,\ref{3} (d) shows the scaled $S_{\infty}$ vs. scaled $v_s$ from the results obtained with two dimensional DEM simulations, where we have also observed a similar spatial segregation. Note that the two dimensional result is not trivial for it is not always obvious that particles produce the same segregation in two and three dimensions \cite{ottino2000mixing}. 

\begin{figure}
\centering
\includegraphics[width=18cm]{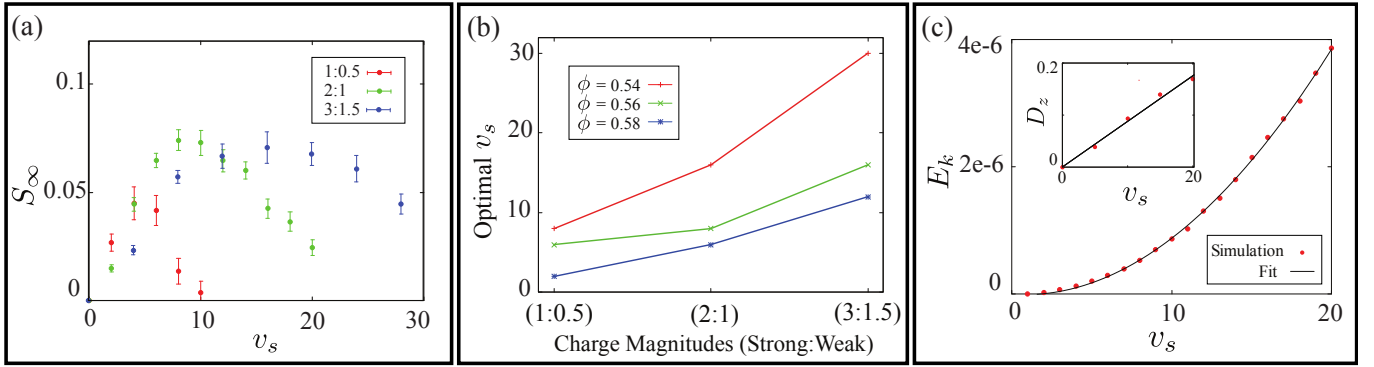}
\caption{(a) $S_{\infty}$ versus $v_s$ for fixed $\phi$ = 0.56 and different charge magnitudes: (strong : weak) = (1:0.5) for red, (2:1) for green, and (3:1.5) for blue. (b) Optimal $v_s$ versus charge magnitudes for several $\phi$. (c) Kinetic energy: $E_k$=$\sum_{i=1}^{N}$ $\frac{1}{2} m_{i} v_{z}^2$, versus $v_s$ From three dimesional simulations  for $\phi$=0.55 (red dots). The black curve is a fit to a quadratic function with correlation coefficient $R^{2}$=0.92. Inset shows diffusion coefficient: $D_{z}$=$<$($z_{i}(t)$-$z_{i}(0)$)$^{2}$$>$/$t$, versus $v_s$ (red dots) also measured in the DEM simulations fitted using a linear function (black line) for $\phi$=0.55. }
\label{4}
\end{figure}

We propose that the maximum segregation results from a competition between mixing and segregational fluxes.  At low shear, particles rearrange in response to electrostatic repulsion, whereas at high shear these rearrangements become overwhelmed by diffusional mixing \cite{chaudhuri2006cohesive, hill1994reversible}. This is essentially the same competition historically analyzed by Einstein \& Sutherland \cite{einstein1905molekularkinetischen, sutherland1905lxxv}. 

Testing this hypothesis is straightforward. We carry out simulations with different charge magnitudes while keeping the ratio between strongly and weakly charged particles unchanged. Increasing the charge magnitudes strengthens the electrostatic repulsion and higher shear is needed to mix particles that rearrange in response to the stronger electrostatic repulsion. Consequently, if the proposed mechanism is correct, increasing the charge magnitude should increase the optimal $v_s$ and vice versa for weaker charges. This is shown in Fig.\,\ref{4} (a), where $S_{\infty}$ is plotted against $v_s$ for different magnitudes of charges. The optimal $v_s$ is increased for a system with higher charges. The same tendency is seen for different packing fractions as shown in Fig.\,\ref{4} (b). Evidently the maximum segregation results from a competition between Coulomb flux and diffusional flux.

In kinetic theory, the diffusion coefficient is expressed as:
\begin{equation}
D_{z}= \mu k_{B} T,
\label{diffusion}
\end{equation}
where $D_z$ is the spanwise diffusivity, $\mu$ is the mobility, $k_{B}$ is Boltzmann's constant, and $T$ is the (granular) temperature, which in our system is the kinetic energy: $T$ = $E_{k}$ = $\sum_{i=1}^{N} \frac{1}{2} m_{i} v_{z}^2$. Notice that we are only interested in the perpendicular motions of particles since the segregation we want to understand is perpendicular to the shearing direction. $\mu$ is a measure of particles' motion through a medium in response to an electric field, and so we use $\mu$ as a surrogate for the segregation for our system: $S$ $\sim$ $\mu$ = $D_{z}$/($k_{B}$$T$). The diffusion coefficient is known to grow linearly with the applied shear rate, which in our case is $v_{s}$ \cite{leighton1987shear} (also see the inset of Fig.\,\ref{4} (c), where the diffusion coefficient is obtained by: $D_{z}$=$<$($z_{i}(t)$-$z_{i}(0)$)$^{2}$$>$/$t$), hence $D_{z}$ $\sim$ $v_s$. For $T$, we find that a suitable fit can be made using a quadratic function with correlation coefficient $R^{2}$=0.92, as shown in the main plot in Fig.\,\ref{4} (c), and so $T$ $\sim$ $a v_{s}^2 + b v_{s} + c$. Consequently, we use a function of the following form to fit the curves in Fig.\,\ref{3} (c):
\begin{equation}
S \sim \frac{v_s}{a v_{s}^2 + b v_{s} + c}.
\label{model}
\end{equation}

The function produces the fit shown in Fig.\,\ref{3} (c) with correlation coefficient $R^{2}$=0.98 for $\phi$ = 0.55. This simple model predicts a maximum as seen in the simulations, but underestimates the decrease in segregation after the maximum. There certainly is a need of additional work to refine our understanding. Evidently from our simple simulations, we find that there is a maximum of segregation as shear grows, and it appears that the competition between the mixing and segregational fluxes can largely account for this maximum. 
 
\section{\label{conclusion}Conclusion}
We found that a binary mixture of similarly charged particles segregate according to their net charge in a simple shear flow. This segregation occurs due to strongly charged particles repelling others and moving outwards to minimize the electrostatic potential energy of the system since the boundaries are insulating. However, we expect the same to happen for conducting boundaries. Our direct simulations showed that for a given packing fraction there is an optimal shear velocity where the segregation maximally intensifies and for a larger shear velocity, diffusion suppresses the segregation. 

\section{\label{acknowledgments} Acknowledgments}
We acknowledge financial support from the European Research Council (ERC) Advanced Grant 319968-FlowCCS and the INCT-SC. NA acknowledges financial support from the Portuguese Foundation for Science and Technology (FCT) under Contract no. UID/FIS/00618/2013, and from the Luso-American Development Foundation (FLAD), FLAD/NSF, Proj. 273/2016. TS acknowledges support from the NSF DMR, award $\sharp$1404792.
\clearpage
\bibliography{bibliography}
\end{document}